# Mini-phoswich and SiPM for Heavy Ion Detection

D. Carbone[a], P. Finocchiaro[a], C. Agodi[a], D. Bonanno[c], D. Bongiovanni[a], M. Cavallaro[a], F. Cappuzzello[a,b], L. Cosentino[a], G. Gallo[b,c], F. Longhitano[c], D. Lo Presti[b,c], S. Reito[c] for the NUMEN collaboration

[a]INFN Laboratori Nazionali del Sud, Catania, Italy
[b]Dipartimento di Fisica e Astronomia, Università di Catania, Italy
[c]INFN Sezione di Catania, Catania, Italy

Abstract
The possibility to use a mini-phoswich detector to identify ions in the region of Z ~ 10 is explored in the framework of the NUMEN project. The NUMEN program, aimed at the investigation of the nuclear matrix elements connected to the neutrinoless double beta decay by means of double charge exchange nuclear reactions, foresees very high fluencies, which prevent the use of standard silicon as stop detectors. The need of reasonable radiation hardness, together with a total energy resolution around 2% and a high granularity, makes scintillators possible candidates. Promising results are obtained using an array of plastic + inorganic phoswich scintillators readout by means of Silicon Photo Multipliers.

Corresponding author: D. Carbone (carboned@lns.infn.it)

## 1. Introduction

One of the most important developments for the technologies implied in nuclear physics is related to the ion detection and identification under extreme conditions of external background and rate. For example, such a detection system is required when a huge ion flux is expected. This is the case of the NUMEN (Nuclear Matrix Elements for Neutrinoless double beta decay) project [1][2], in which the foreseen counting rate is about 50 kHz/cm$^2$. The aim of the project is to access the nuclear matrix elements entering the expression of the life-time of neutrinoless double beta decay (0νββ) by cross section measurements of heavy-ion induced double charge exchange reactions. In particular, the ($^{20}$Ne,$^{20}$O) and ($^{18}$O,$^{18}$Ne) reactions are measured at incident energies ranging from 10 to 60 MeV/amu. The use of the MAGNEX large acceptance spectrometer [3,4] to detect the ejectiles produced in the nuclear collisions allows to clearly identify the ions of interest and to extract the absolute cross section for the ground-state-to-ground-state transition with high accuracy, as demonstrated in a pilot experiment on $^{40}$Ca$_{g.s.}$($^{18}$O,$^{18}$Ne)$^{40}$Ar$_{g.s.}$ [5].

Due to the very low cross section (of the order of few nb) involved in the reactions of interest for NUMEN, a substantial upgrade of the INFN-LNS Superconducting Cyclotron (CS) [6] and beam lines is foreseen to reach two orders of magnitude more beam current. On the side of the detection system, however, the present setup is not reliable for the high rate measurements required.

The main limitations come from the present MAGNEX Focal Plane Detector (FPD) [7], which consists of a wire-based gas tracker, working as a drift chamber, and a wall of stopping silicon pad detectors. The sustainable rate of the MAGNEX FPD is ~ 5 kHz, due to the slow drift of positive ions from the multiplication wires to the Frisch grid. To overcome this limit, the multiplication wires will be replaced by a series of micropatterned electron multiplier foils. This technology is not able to provide accurate information on the ion energy loss, thus such an issue must be demanded to the wall of stopping telescope detectors downstream the tracker. The present silicon stopping detectors are not feasible for the high detection rate: they can only operate up to fluences ≤ 10$^9$ implanted ions/cm$^2$ (for heavy ions) whereas the fluence expected in ten years of the NUMEN research program is ~ 10$^{14}$ ions/cm$^2$.

In addition to the requirement of radiation hardness, additional constraints must be taken into account for the design of the new stopping wall telescope for particle identification: i) energy resolution better than 2% to keep the same particle identification performance of the current FPD (~ 1/160 mass resolving power [8]); ii) time resolution of the order of 1÷ 2 ns in order to guarantee an accurate measurement of the time of flight of the gamma rays detected by the foreseen calorimeter mounted at the target position [9]; iii) high segmentation (modules of 1 cm$^2$) to limit the double-hit probability below 10% in the whole FPD. A promising candidate for the new stop detector could be Silicon Carbide semiconductor detectors, which are still under development [10]. Another possibility is the use of an array of plastic + inorganic *phoswich* scintillators [12] readout by means of Silicon Photo Multipliers (SiPM), which will be discussed in the following. A phoswich detector (also called *phosphor sandwich*) is the combination of two dissimilar scintillators chosen to have different decay times optically coupled to a single Photo Multiplier tube [11]. In this way, the shape of the output pulse from the SiPM is dependent on the relative contribution of scintillation light from the two scintillators.

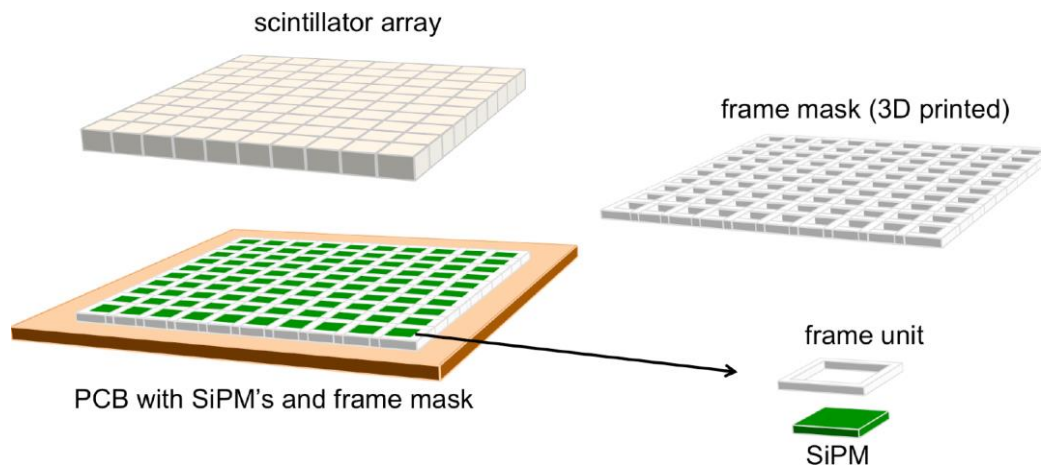

Fig. 1. Pictorial view of the components of the detector array.

## 2. Experimental setup and results

### 2.1. The prototype

A square array of 100 phoswich detectors was built as a prototype demonstrator of a feasible stopping wall. Each element has 1 cm x 1 cm area and is made up of a 200 μm thick fast plastic scintillator (Pilot-U) (decay time 1.8 ns) followed by a 5 mm thick CsI(Tl) inorganic scintillator (decay time 3 μs). The scintillators are embedded in a white epoxy frame which acts as mechanical support for the matrix and as optical reflector. The fast plastic scintillator, in the current version, was placed on top of the matrix as a single sheet optically coupled to it by means of a very thin layer of silicone grease. The external face of the plastic was covered with a 50 μm thick aluminized Mylar foil. The scintillation light is readout by means of 100 Silicon Photo Multipliers (SiPM), 6 mm x 6 mm active area produced by SensL [13], installed on a dedicated PCB board designed for this purpose and featuring a single connector for the voltage bias and eight 13-pin dual-in-line connectors for the individual signal output of each detector cell. A white polymeric frame was 3D-printed to mechanically host and align the detector matrix to the SiPM array (Fig.1). Moreover, this frame also acts as additional reflector covering the spare area of each CsI(Tl) scintillator around its SiPM. The optical coupling between SiPM and scintillator was done by means of the previously mentioned silicone grease. A picture of the assembled prototype is shown in Fig.2.
Taking as reference case for NUMEN an $^{16}$O at 20 MeV/amu, the expected energy loss in the ΔE stage is ~37 MeV, and the thickness of the CsI(Tl) scintillator is sufficient to stop the ejectile.

## 2.2. Experimental results

An in beam test of 1 cm x 1 cm single phoswich was performed with a 46 MeV $^7$Li beam, delivered by the Tandem accelerator of INFN-LNS, on al $^7$LiF + C target. From a rough analysis of the data the resulting resolution for the elastic peak was found to be ~ 2.4% [14]. The same prototype was tested also with a 320 MeV $^{16}$O beam, accelerated by the Superconducting Cyclotron of INFN-LNS, on a $^{27}$Al target. The shape of the output signal from the phoswich detector depends on the relative contribution of scintillation light from the fast plastic scintillator (Pilot-U) called *fast component* and the *slow component* from the second stage of CsI scintillator. A BafPro filter & amplifier module [15][16], designed by INFN-Milano, was used to replicate the detector signal in two copies, filtering out from one copy the fast light and amplifying both. Removing the fast light component was found to be more effective than just selecting it.

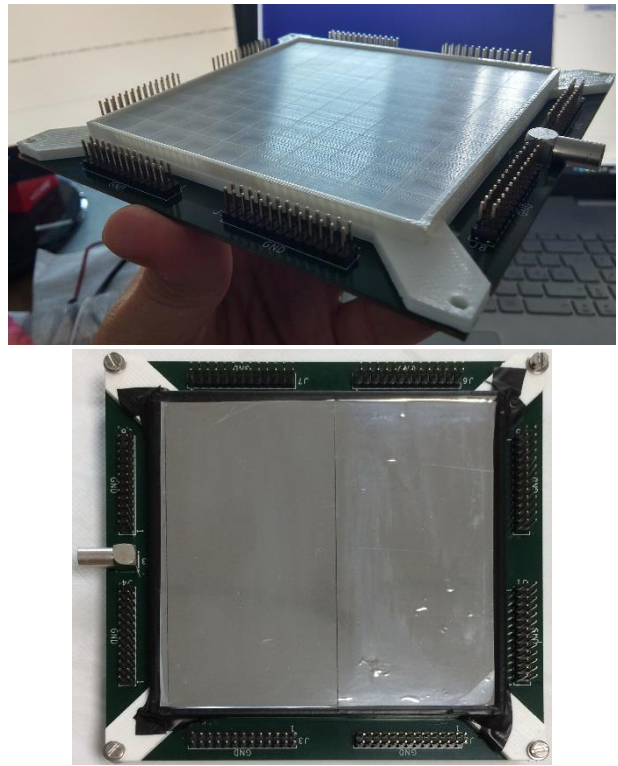

Fig. 2. Side view (top) and top view (bottom) of the assembled 100-phoswich prototype.

The scatter plot of the total light (X-axis) versus the slow light (Y-axis) is shown in Fig.3 (upper panel). In this representation, the discrimination capability of the phoswich detector is evident. Indeed, for a fixed value of the slow component (energy released in the CsI scintillator), the higher Z ions have a larger total light because they lose higher energy in the Pilot-U stage. Moreover, the quenching effect for the light generated by the higher Z ions decrease the slow component, thus allowing a larger separation between the high and low Z values. In Fig.3 (upper panel) the different loci correspond to different atomic number Z (and possibly different mass) and the separation is better for higher Z values. In order to evaluate the discrimination capability of this technique we selected a slice of ~ 500 channels (as indicated by the black lines in Fig. 3 upper panel) on the Total light and projected it into a one-dimensional spectrum of Slow* obtained by a transformation, Slow* = Slow – f(Total), which rectifies the branches and optimizes the width of the projected peaks of the different ions over the new vertical axis. The projection result is shown in the lower panel of Fig.3. A few structures are visible, corresponding to Z = 2 (α), Z = 1 in which the proton (p) locus is well separated from triton (t) and deuteron (d), and Z = 0 (γ). The distance between the centroids of the α and p structures corresponds to 15 σ units (p) and 12 σ units (α).

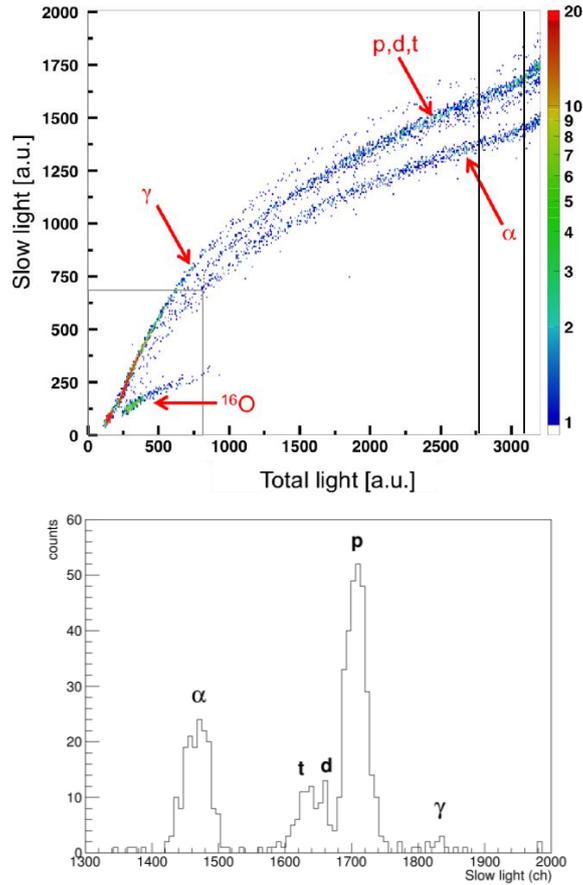

Fig. 3. Upper panel: slow light versus total light for test with $^{16}$O at 320 MeV incident energy on $^{27}$Al target, the separated loci correspond to γ, (p, d, t), α and $^{16}$O. Lower panel: Projection on the slow axis of the upper panel with a selection on the total light axis as indicated by the black lines in the upper panel.

In order to investigate the region of higher Z, an amplification factor of 4 was applied to both components and the result is shown in Fig.4 (upper panel). We selected a slice of ~ 400 channels (as indicated by the black lines in Fig.4 upper panel) on the Total light and projected into a one-dimensional spectrum of Slow* obtained by a similar transformation as the previous one. The projection result is shown in the lower panel of Fig.4. The region projected in the one-dimensional spectrum corresponds to an energy loss of ~ 100 MeV for the impinging $^{16}$O. The obtained Full Width at Half Maximum (FWHM) is ~ 2.8 %, that extrapolated to the NUMEN reference case corresponds to $2.8\% \sqrt{\frac{100 \text{MeV}}{300 \text{ MeV}}} \sim 1.6\%$, thus slightly better than the required value (~ 2%).

A preliminary test of the new 100-phoswich prototype was also performed with $^{137}$Cs and $^{22}$Na radioactive sources. The $^{137}$Cs source produces gamma rays at 662 keV, whereas the $^{22}$Na source emits positrons which annihilate producing 511 keV and also 1274 keV gamma rays. The signal from the SiPM was preamplified x10 by means of a Phillips Scientific 776 and further amplified by means of an ORTEC 572 spectroscopy amplifier (1μs shaping time), whose output was sent to a multichannel analyser. A satisfactory uniformity of response was found among all the channels. Two spectra from one channel, with their characteristic shapes, are shown in Figs. 5 and 6. It has to be remarked that the obtained energy resolution with gamma rays is not optimal, due to a worsening introduced by a not perfect optical coupling between the thin plastic scintillator and the CsI(Tl). This will be improved in the forthcoming tests.

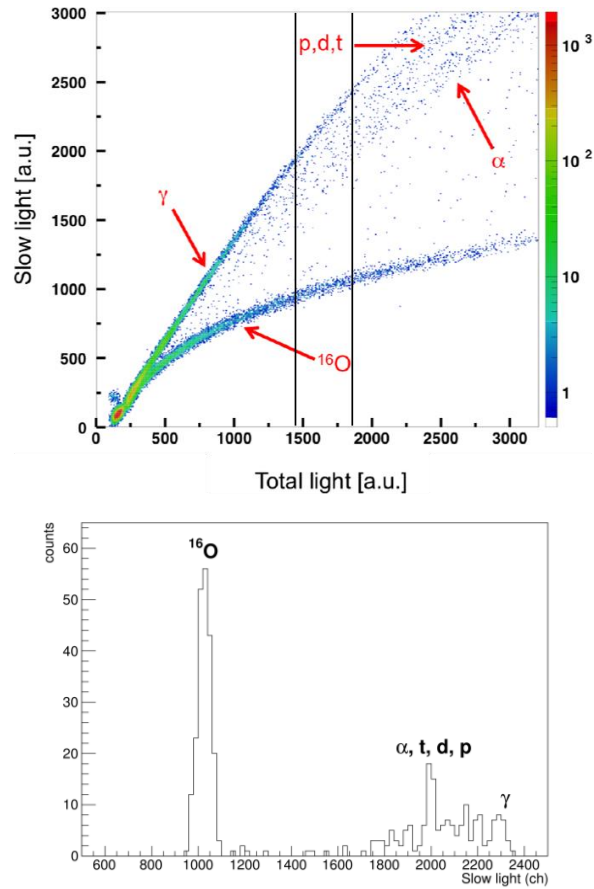

Fig. 4. Upper panel: slow light versus total light for test with $^{16}$O at 320 MeV incident energy on $^{27}$Al target with a 4X amplification on both components with respect to the upper panel of Fig.3. The separated loci correspond to γ, (p, d, t), α and $^{16}$O. Lower panel: Projection on the slow axis of the upper panel with a selection on the total light axis as indicated by the black lines in the upper panel.

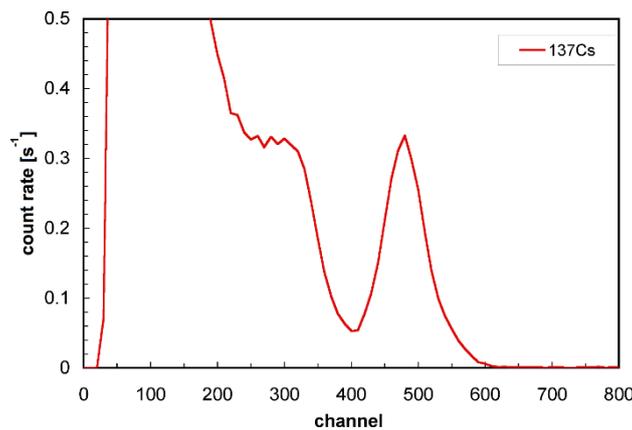

Fig. 5. Energy spectrum measured with a $^{137}$Cs source by one channel of the phoswich array.

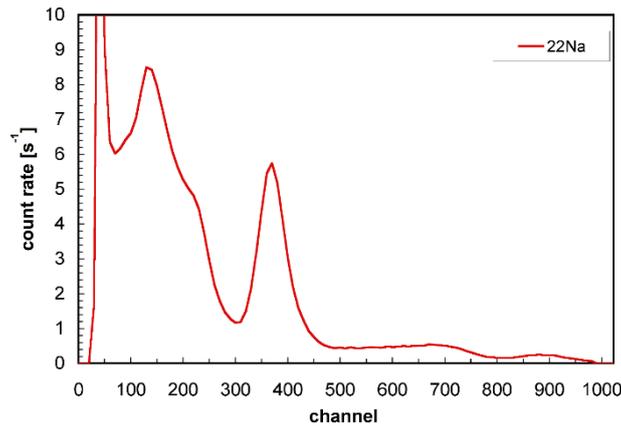

Fig. 6. Energy spectrum measured with a $^{22}$Na source by one channel of the phoswich array.

## 3. Conclusions

The possibility to use an array of mini-phoswich detectors readout by means of Silicon Photo Multipliers to identify heavy ions is explored. In particular, the energy resolution performances are evaluated taking into account the requirements of the NUMEN project (resolution better than ~ 2%). In beam tests have been performed using $^7$Li at 46 MeV incident energy and $^{16}$O at 320 MeV as well as radioactive sources measurements. The obtained resolutions are promising since a value of 1.6 % has been obtained from an extrapolation to the NUMEN reference case.
The very compact, robust and inexpensive phoswich detector promises to be a realistic candidate for the NUMEN project and also for other applications needing the detection and identification of heavy ions.

## Acknowledgments

This project has received funding from the European Research Council (ERC) under the European Union's Horizon 2020 research and innovation programme (grant agreement No 714625).